# AN ADAPTIVE APPROACH TO FILTER A TIME SERIES DATA

[1]Koushik Ghosh and [2]Probhas Raychaudhuri

[1]Department of Mathematics
University Institute of Technology
University of Burdwan
Burdwan-713 104
INDIA
Email: koushikg123@yahoo.co.uk

[2]Department of Applied Mathematics
University of Calcutta
92, A.P.C. Road, Kolkata-700 009
INDIA
Email: probhasprc@rediffmail.com

**ABSTRACT:**

A physical data (such as astrophysical, geophysical, meteorological etc.) may appear as an output of an experiment or it may come out as a signal from a dynamical system or it may contain some sociological, economic or biological information. Whatever be the source of a time series data some amount of noise is always expected to be embedded in it. Analysis of such data in presence of noise may often fail to give accurate information. The method of filtering a time series data is a tool to clean these errors as possible as we can just to make the data compatible for further analysis. Here we made an attempt to develop an adaptive approach of filtering a time series and we have shown analytically that the present model can fight against the propagation of error and can maintain the positional importance in the time series very efficiently.

**INTRODUCTION:**

Analysis of data is a very important task since it is the source of information. It provides the validation of theories and models as well as their improvements. Data analysis sometimes can give birth of a new theory or model. But in reality the problem is that any kind of time series data either from an experiment or from a dynamical system, or from any economic, sociological or biological aspect, usually contains systemic or manual error. Analysis of such data in presence of noise often leads to a wrong interpretation of the data. So we need to develop an initial platform by denoising the data from which we can start the extensive study on it. Filtering a time series data is always an indispensable task to deal with.
We have a number of existing methods for filtering a time series data. We have traditional 3-point or 5-point moving average method as an initial technique to smooth the data. We have Empirical Mode Decomposition (Low Pass Filtering, High Pass Filtering), Monte Carlo Technique and Wavelet Analysis to serve this particular purpose.

In this list we also have very well known Kalman Filter [1, 2] and Simple Exponential Smoothing [3, 4, 5].

While concentrating on the process of reducing the noise we must remember one thing that we cannot compromise heavily with the trend and characteristic of the time series data. But it is violated in the case of 3-point and 5-point moving average methods. If the original time series be $\{x_i\}$; i=1, 2, 3… n then in 3-point moving average the processed data at the i-th position is given by $y_i = (x_{i-1}+x_i+x_{i+1})/3$; i=2, 3, …., (n-1) and in 5-point moving average the same is $y_i=(x_{i-2}+x_{i-1}+x_i+x_{i+1}+x_{i+2})/5$; i=3, 4,, …., (n-2). This clearly reveals that we are missing proper information at i-th place as we are busy with making average of the information at the neighbouring places instead of giving proper weight to the information at i-th place of the original time series. Moreover, we miss two vital data in 3-point moving average method. The worst situation occurs in the case of 5-point moving average method; here we miss four important data. In the case of Empirical Mode Decomposition (Low Pass or High Pass Filter) we rely upon all the local extrema of the raw data. In that case we start the procedure by connecting all the local maxima by a cubic spline to form an upper envelope and repeat the procedure for the local minima to form a lower envelope [6]. So to identify all the local extrema we first have to plot the original data against time. From the plot it is difficult to identify the local extrema accurately. So, there is always a high chance to start the operation with a handful of error. In Monte Carlo Technique the process of denoising the data is performed by the method of random number generation assuming the noise present in the data to be Gaussian. But this kind of treatment often fails in reality since the real world problems don't always play by the rules. In Wavelet Analysis we adopt the technique of compression of data. But it is well understood that in the process of compression of data the noise with a very high magnitude present at any position of a time series usually reduces its magnitude by distributing that deleted part of the noise to the other positions. Furthermore, the time series data on application of this method shows higher magnitudes at the tail. This can certainly generate wrong interpretations of the data.

Conceptually Kalman filter of a time series data $\{x_i\}$; i=1, 2, … , n is analogous to updating a running average [1, 2, 7]:

$$y_i=\{(i-1)/i\}x_{i-1}+(1/i)x_i; i=1, 2, …, n.$$

It should work in far better manner than the 3-point or 5-point moving average method as in the search of the i-th information this method concentrates only on the (i-1)-th and i-th data of the original time series by taking their convex linear combination rather than by starting over each time with simple method of average.

In the case of Simple Exponential Smoothing the prescribed model for a time series data $\{x_i\}$; i=1, 2, …., n is [3, 4, 5] $y_1=x_1$ and $y_i=\alpha x_i+(1-\alpha)y_{i-1}$; i=2, 3, …., n where $y_i$ is the smoothed data at the i-th position and $\alpha$ ($0< \alpha< 1$) is a parameter. This is analogous to $y_1=x_1$ and $y_i=\alpha x_i+\alpha(1-\alpha)x_{i-1}+\alpha(1-\alpha)^2 x_{i-2}+….+\alpha(1-\alpha)^{i-2}x_2+(1-\alpha)^{i-1}x_1$ for i=2, 3, …, n where the sum of the corresponding weights $\alpha$, $\alpha(1-\alpha)$, $\alpha(1-\alpha)^2$, …., $\alpha(1-\alpha)^{i-2}$ and $(1-\alpha)^{i-1}$ is equal to unity. Thus in effect, each smoothed value is a convex linear combination of all the previous observations as well as the current observation.

We believe that denoising a time series data can be performed with a satisfactory level of accuracy if we concentrate on the following two matters:

i) Certain cases may arise in which the error generated in a certain position propagates to the next stages. In those cases while trying to develop a smoothing model we must consider this propagation of error and must try to fight against it.
ii) While extracting the new time series data by filtering the old one we must keep in mind the positional importance of data i.e. if $\{y_i\}$ be the newly developed time series data by filtering the old one $\{x_i\}$; i=1, 2, …., n the $y_i$'s must be generated mostly from the corresponding $x_i$'s. In the case of Kalman filter the positional importance is not maintained since in the expression of i-th filtered data more weightage is given in $x_{i-1}$ than in $x_i$.

Keeping the above two points in mind in this study we make an effort to develop an adaptive model of filtering of a time series data.

**THEORY: FORMULATION OF GENERAL ADAPTIVE RULE OF FILTERING:**

We take a finite time series data $\{x_i\}$; i=1,2, …., n. It is expected that the time series data contains some noise. We assume that $\{x_i^{/}\}$; i=1, 2, …., n is the actual time series in the absence of noise. We concentrate in those situations where the observed data at any position depends on the observed results at the preceding positions. In these situations if the technique of collecting the data is not sound enough there is always a chance that the error terms may propagate to their next stages. In those cases we can write:

$$x_i = x_i^{/} + \sum_{j=1}^{i-1} a_{ij}\varepsilon_j + \varepsilon_i; \quad i=1, 2, …., n \tag{1}$$

Here $\varepsilon_i$ is the error generated at the i-th observation; $\varepsilon_j$'s [j=1, 2,, …., (i-1)] are the errors at the preceding stages propagated to the i-th step with factors $a_{ij}$ associated to them. We next particularize our case by believing that in the process at each stage the effects of preceding errors get enlarged and therefore we have $a_{ij}$'s as corresponding enlarging factors, where each $a_{ij}>1$ and $a_{kj}$'s are arranged in ascending error for a fixed j [j=1, 2, …., (n-1); k=j+1, j+2, …., n].

Next we introduce certain weights in the time series data in the following manner:

$$y_i = \sum_{k=1}^{i} w_{ki} x_k; \quad i=1, 2, …, n \tag{2}$$

in which all $w_{ki} \varepsilon (0, 1)$.

We restrict ourselves by taking a convex linear combination in (2) and therefore

$$\sum_{k=1}^{i} w_{ki} = 1 \tag{3}$$

Using (1) in (2) we get

$$y_i = \sum_{k=1}^{i} w_{ki} x_k^{/} + \sum_{k=1}^{i} w_{ki} (\varepsilon_k + \sum_{j=1}^{k-1} a_{kj} \varepsilon_j); \quad i=1, 2, \ldots, n \tag{4}$$

We must remember one thing that in order to maintain the positional importance at each stage for a fixed i, $w_{ki}$ must increase steadily with the increase in k so that it can attain its maximum value at k=I and the maximum weight $w_{ii}$ can be associated with $x_i^{/}$. Side by side, these weights are to be taken in such a manner so that the expression for error at each stage can be reduced effectively in the process.

If there is no propagation of error i.e. if all $a_{ij}=0$ and $\varepsilon_i$'s are arranged in ascending order we recommend the same technique of filtering as stated above.

We cite one example of choosing the weight. We take

$$w_{ki} = (i)^{k-1} / \sum_{l=1}^{i} (i)^{l-1}; \quad i=1, 2, \ldots, n; \quad k=1, 2, \ldots, i \tag{5}$$

so that

$$\sum_{k=1}^{i} w_{ki} = \sum_{k=1}^{i} (i)^{k-1} / \sum_{l=1}^{i} (i)^{l-1} = 1 \tag{6}$$

Using (5) in (4) we have,

$$y_i = [\sum_{l=1}^{i} (i)^{l-1}]^{-1} [\sum_{k=1}^{i} (i)^{k-1} x_k^{/} + \sum_{k=1}^{i} (i)^{k-1} (\varepsilon_k + \sum_{j=1}^{k-1} a_{kj} \varepsilon_j)]; \quad i=1, 2, \ldots \tag{7}$$

It can be easily noticed that the positional importance in this case is maintained at each stage since our present $w_{ki}$ has a monotonic increasing profile with the increase in k so that the expression $(i)^{k-1} / \sum_{l=1}^{i} (i)^{l-1}$ attains its maximum value at k=i and the maximum weight $(i)^{i-1} / \sum_{l=1}^{i} (i)^{l-1}$ is associated with $x_i^{/}$. Initially in the original time series the error term involved in the i-th data was $\sum_{j=1}^{i-1} a_{ij} \varepsilon_j + \varepsilon_i$. Applying the filter as shown in the error becomes $[\sum_{l=1}^{i} (i)^{l-1}]^{-1} \sum_{k=1}^{i} [(i)^{k-1} (\varepsilon_k + \sum_{j=1}^{k-1} a_{kj} \varepsilon_j)]$. Since each $a_{ij}>1$ and $a_{kj}$'s are arranged in ascending order for a fixed j [j=1, 2, …., (n-1); k=j+1, j+2, …., n] it can be easily observed that error generated in (7) after applying the present filter is significantly less in magnitude than the error present in (1) for actual data at each i. We can experience a similar kind of observation for a time series data where all $a_{ij}=0$ and $\varepsilon_i$'s are arranged in ascending order.

We also have certain examples where the process of observation learns from previous experience and as time progresses the system gets more and more efficient to tackle the error. In those cases, any kind of propagation of error is not expected and $\varepsilon_i$'s should be arranged in descending order. For those particular situations, we write down the time series data in reverse order i.e. we form a new time series $\{x_i^*\}$; i=1, 2, …., n in which $x_i^* = x_{n-i+1}$. In this time series data the corresponding error terms will be in ascending order. We suggest to apply the same model of filtering on $\{x_i^*\}$. After having the filtered data from $\{x_i^*\}$ we again write down them in reverse order to get the final smoothed time series.

## SIMPLE EXPONENTIAL SMOOTHING FROM THE PERSPECTIVE OF OUR PRESENT MODEL:

In Simple Exponential Smoothing we rewrite the expression for the i-th smoothed data in the following manner:
$y_1 = x_1$
and,
$$y_i = (1-\alpha)^{i-1}x_1 + \alpha(1-\alpha)^{i-2}x_2 + \ldots + \alpha(1-\alpha)^2 x_{i-2} + \alpha(1-\alpha)x_{i-1} + \alpha x_i \text{ for } i=1, 2, 3, \ldots, n \tag{8}$$

In order to maintain the positional importance in this smoothing we must have the weights in (8) in ascending order and it is possible only when $\alpha > 0.5$.

On application of this smoothing the error term at i-th stage becomes

$$[(1-\alpha)^{i-1} + \alpha(1-\alpha)^{i-2}a_{21} + \ldots + \alpha(1-\alpha)^2 a_{i-2,2} + \alpha(1-\alpha)a_{i-1,1} + \alpha a_{i1}]\varepsilon_1 + \sum_{j=2}^{i-1}[\alpha(1-\alpha)^{i-j-1} + \alpha(1-\alpha)^{i-j-2}a_{j+2,j} + \ldots + \alpha(1-\alpha)a_{i-1,j} + \alpha a_{ij}] + \alpha\varepsilon_i,$$

while in the original time series data the error term present in i-th data is $\sum_{j=1}^{i-1} a_{ij}\varepsilon_j + \varepsilon_i$. If we believe that each $a_{ij} > 1$ and $a_{kj}$'s are arranged in ascending order for a fixed j [j=1, 2, …., (n-1); k=j+1, j+2, …., n] it can be easily seen that the error generated after employing the present smoothing is less in magnitude than the error present in the actual data for each i. Similar result can be observed for a time series data in which all $a_{ij}=0$ and $\varepsilon_i$'s are arranged in ascending order. Again, if for a time series data all $a_{ij}=0$ and $\varepsilon_i$'s are arranged in descending order we can write down the time series data in reverse order and proceed as earlier to have the similar conclusion. Thus we can certainly accept the Simple Exponential Smoothing as a favourable example of our General Adaptive Rule of Filtering. In absence of the propagation of error the percentage reduction of the total error present in the data by the application of Simple Exponential Smoothing is given by

$$\{100/(\sum_{i=1}^{n}\varepsilon_i)\}[\{(1-\alpha)^n - (1-\alpha)\}(\varepsilon_1/\alpha) + \sum_{i=1}^{n-1}(1-\alpha)^{n-i}\varepsilon_{i+1}]\%.$$

For sufficiently large n in order to have a sizable amount of reduction in the total error we must have $\alpha$ in the right hand side neighbourhood of 0.5.

## CONCLUSION:

Our present analysis strongly recommends that in Simple Exponential Smoothing when the error terms are in ascending or descending error we must have $0.5<\alpha<1$. Although text book data filtering theory is primarily concerned with the presence of random errors, but in reality, errors in data are often systematic rather than random [1]. If we still go for a situation where the error terms are not in steadily ascending or descending order we should not feel ourselves helpless. At present we suggest to make the smoothing first by assuming the errors in ascending order and then to do the same assuming the errors in descending order and finally by making the simple average of these two outcomes we can expect to get the desired smoothed data. If we want to take the help of Simple Exponential Smoothing to serve this particular purpose just for the sake of positional importance we must have $(1/3)<\alpha<1$. Consequently we can expect a successful end of the controversy regarding the range of $\alpha$ to be assumed. If it is felt that the magnitude of the error present in a time series data is very high we suggest applying our present rule of adaptive filtering repeatedly for several numbers of times.

Our General Adaptive Rule of Filtering has reliance on linearity. An extension of our present model of filtering in nonlinear format can be a very interesting field of future study.